\documentclass[english,aps, superscriptaddress, floatfix, 12pt]{revtex4}
\usepackage[T1]{fontenc}
\usepackage[latin1]{inputenc}
\usepackage{graphicx}
\usepackage{amssymb}
\usepackage{epsfig}
\usepackage{amsmath}
\usepackage{psfrag}

\usepackage{babel}
\makeatother
\begin{document}

\title{Contributions of stochastic background fields to the shear and bulk viscosities of the gluon plasma}

\author{Dmitri Antonov\\
{\it Departamento de F\'isica and Centro de F\'isica das Interac\c{c}\~oes Fundamentais,}\\ 
{\it Instituto Superior T\'ecnico, UT Lisboa,
Av. Rovisco Pais, 1049-001 Lisboa, Portugal}}

\noaffiliation

\begin{abstract}
The contributions of confining as well as nonconfining nonperturbative self-interactions of 
stochastic 
background fields to the shear and bulk viscosities of the gluon plasma in SU(3) Yang--Mills theory are calculated. 
The nonconfining self-interactions change (specifically, 
diminish) the values of the shear and bulk viscosities by 15\%, that is close to the 17\% which the 
strength of the nonconfining self-interactions amounts of the full strength of nonperturbative 
self-interactions.
The ratios to the entropy density of the 
obtained nonperturbative contributions to the shear and 
bulk viscosities are compared with the 
results of perturbation theory and the predictions of ${\cal N}=4$ SYM. 
\end{abstract}

\maketitle

\section{Introduction}
Ultrarelativistic nucleus-nucleus collisions performed at RHIC indicate that, in the vicinity 
of the deconfinement critical temperature, the quark-gluon plasma 
can behave more like liquid than a dilute gas of quarks and gluons.
One indication of this kind is the experimentally found 
common velocity of different species of particles, which are 
emitted from the expanding fireball of the quark-gluon plasma.  
The corresponding experimental results~\cite{odin, dva} 
can be successfully described by the relativistic hydrodynamics of an almost non-viscous liquid~\cite{khhet}.
The energy-momentum tensor of an ideal (i.e. absolutely non-viscous) liquid 
has the form~\cite{LL} $\Theta_{\mu\nu}=-pg_{\mu\nu}+Tsu_\mu u_\nu$,
where $s=s(T)$ is the entropy density, and $u_\mu$ is the velocity of energy transport.
The principal deviation from the ideality is described by 
a correcting term $\Delta \Theta_{\mu\nu}$, which 
depends on the derivatives of the velocity linearly:
$$\Delta \Theta_{\mu\nu}=\eta_T\cdot(\Delta_\mu u_\nu+\Delta_\nu u_\mu)+
\left(\frac23\eta_T-\zeta_T\right)H_{\mu\nu}\partial_\rho u_\rho,$$
where $H_{\mu\nu}=u_\mu u_\nu-g_{\mu\nu}$, $\Delta_\mu=\partial_\mu-u_\mu u_\nu\partial_\nu$.
The temperature-dependent coefficients, $\eta_T$ and $\zeta_T$, are called 
respectively the shear and the bulk viscosities. Together, they are called first-order transport 
coefficients of the liquid. 

In hydrodynamics~\cite{LL}, the shear viscosity characterizes a change in shape of a 
fixed volume of the liquid, whereas the bulk viscosity characterizes a change 
in volume of the liquid of a fixed shape. 
In particular, the bulk viscosity enters the Navier--Stokes' equation through the term 
$\left(\zeta_T+\frac13\eta_T\right){\,}{\rm grad}{\,}{\rm div}{\,}{\bf v}$, being therefore relevant only 
when ${\rm div}{\,}{\bf v}\ne 0$, i.e. for compressible liquids. 
The shear viscosity enters the Navier--Stokes' equation foremost through the term
$\eta_T{\,}\Delta{\bf v}$, representing the ability of particles to transport momentum.
Experimental data for water, helium, and nitrogen, analyzed recently in Ref.~\cite{ckm}, show the minimum 
of the shear-viscosity to the entropy-density ratio ($\eta_T/s$) and the maximum of the 
bulk-viscosity to the entropy-density ratio ($\zeta_T/s$) in the vicinity of the liquid-gas phase-transition 
critical temperature.  
Given different types of liquid-gas phase 
transitions and different types of molecules for these three substances, one can guess 
that the temperature-behaviors of $\eta_T/s$ and $\zeta_T/s$ are universal, and thus can be qualitatively the same also for the quark-gluon plasma.

Furthermore, it is worth noticing that the ($\eta_T/s$)-ratio has been calculated in 
${\cal N}=4$ SYM theory~\cite{pss},
where it equals to the temperature-independent constant 
$1/(4\pi)$. [Note that, since ${\cal N}=4$ SYM theory is a conformal field theory, 
its deconfinement phase can start only right 
at $T=0$, for which reason one should not expect any temperature-dependence of the ($\eta_T/s$)-ratio.]
This constant is at least by one order of magnitude smaller than the minima 
of the above-mentioned empirical results for water, helium, and nitrogen, as well as the values of the 
($\eta_T/s$)-ratio in perturbative QCD. 
Such a smallness of this ratio in ${\cal N}=4$ SYM has led to the widely known conjecture that the 
finite-temperature version of ${\cal N}=4$ SYM provides (a theoretic example of) the most perfect quantum liquid. For example, at the temperature of 200~MeV, the length of a mean free path of a 
parton traversing such a liquid is as small as $\lesssim$0.1~fm. However, since ${\cal N}=4$ SYM 
is a conformal field theory, the bulk viscosity $\zeta_T$ in this theory is strictly zero.
That makes ${\cal N}=4$ SYM different from the real QCD, where 
the effects of non-conformality are essential (e.g. in the so-called interaction measure $\varepsilon-3p$) up to the temperatures as high as $2\div3$ times the deconfinement critical temperature $T_c$.

In a quantum field theory, the contributions of 
various fields to the viscosities can be calculated by means of the Kubo formulae.
A Kubo formula is an integral equation,
which expresses the spectral density $\rho_T(\omega)$ of a given viscosity 
via the corresponding finite-temperature two-point Euclidean correlation function 
of the energy-momentum tensor~\cite{kw}, specifically 
$\bigl<\Theta_{12}(x)\Theta_{12}(y)\bigr>_T$ for the shear viscosity and 
$\bigl<\Theta_{\mu\mu}(x)\Theta_{\nu\nu}(y)\bigr>_T$ for the bulk viscosity. 
In particular, the fact that it is the trace anomaly $\Theta_{\mu\mu}(x)$, which is correlated in case 
of the bulk viscosity, explains in terms of the Kubo formula why the bulk viscosity vanishes in any conformal field theory. Below, we study a more realistic case of SU(3) Yang--Mills theory. There, the energy-momentum tensor can receive contributions from the stochastic background fields and the so-called valence gluons. 
The low-energy nonperturbative stochastic background fields are characterized  
by the temperature-dependent chromomagnetic gluon condensate $\bigl<g^2(F_{ij}^a)^2\bigr>_T$ and the 
correlation length of the chromomagnetic vacuum, $1/\mu_T$, that is the distance at which the 
two-point correlation function of the chromomagnetic fields exponentially falls off~\cite{ds, dmp}.
The valence gluons are higher in energy than the background fields, and are confined by these fields 
at large {\it spatial} separations~\cite{s}. This property of the valence gluons makes them different 
from the ordinary gluons of perturbation theory. 
Such a two-component model of the gluon plasma was recently proven efficient (cf. Refs.~\cite{yus}, \cite{yus1}) for the description of the pressure and the interaction measure, which had been simulated on the lattice in Ref.~\cite{f1}, as well as for the calculation of the radiative energy loss of a highly energetic parton traversing the plasma~\cite{epj}.

It therefore looks natural to apply the same two-component model of the gluon plasma to a derivation 
of the shear and bulk viscosities by means of the Kubo formulae. The first step in this 
direction was made in Ref.~\cite{1}, where 
a contribution of {\it confining} self-interactions of the stochastic background fields to the shear viscosity was found. Besides those, lattice simulations~\cite{dmp, na, latt} point to the existence 
of also {\it nonconfining} nonperturbative self-interactions of the background fields, whose 
strength amounts to 17\% of the full strength of nonperturbative self-interactions. In this paper, we 
calculate the contribution to the shear viscosity produced by  
nonconfining nonperturbative self-interactions of the background fields, 
and show that it diminishes the contribution of confining 
self-interactions by 15\% (that is rather close to the above-mentioned 17\%). We furthermore calculate for the first time the contributions 
of both confining and nonconfining nonperturbative self-interactions to the bulk viscosity of the gluon plasma.

Note that the Kubo formulae lead to the following temperature dependence of the contributions 
of the background fields to the viscosities~\cite{1}:
$$\eta_T\propto\zeta_T\propto\frac{\bigl<g^2(F_{ij}^a)^2\bigr>_T^2}{\mu_T^5}.$$  
At temperatures above that of the dimensional reduction, $T>T_{*}$ (where $T_{*}$ lies somewhere 
between $T_c$ and $2T_c$~\cite{f1}), one has $\eta_T\propto\zeta_T\propto (g_T^2T)^3$, whereas
the temperature dependence of the entropy density at $T\gtrsim 2T_c$ is $s\propto T^3$. Therefore, 
at such rather high temperatures, 
one expects that the ratios to the entropy density of the contributions of the background fields to 
the two viscosities exhibit the same temperature behavior:
$$\frac{\eta_T}{s}\propto\frac{\zeta_T}{s}\propto g_T^6~~~ {\rm at}~~~ T\gtrsim 2T_c.$$
In this paper, we prove that this is indeed the case, and explicitly calculate numerical coefficients 
in the formulae above. The contributions of valence gluons to the shear and bulk viscosities 
will be addressed in the future publications.

The paper is organized as follows. In the next section, we first review our approach, 
and then use it to calculate  
the contribution of confining self-interactions of the background fields to the bulk viscosity of the gluon plasma, $\zeta_T$. In section~III, we find contributions to both  
$\eta_T$ and $\zeta_T$ produced by nonconfining nonperturbative self-interactions of the background fields.
In section~IV, we perform a numerical evaluation of the obtained contributions 
to the viscosities, and compare their ratios to the entropy density with those known from other approaches.
In section~V, we summarize the results of the paper.

\section{Contributions of confining self-interactions of the background fields}
We start with recollecting the approach to the calculation of the shear viscosity $\eta_T$, suggested in Ref.~\cite{1}, and generalizing it to the calculation of the bulk viscosity $\zeta_T$. The two viscosities
can be defined through their spectral densities, $\rho^{(s)}_T(\omega)$ and 
$\rho^{(b)}_T(\omega)$, as~\cite{kw}
\begin{equation}
\label{vb53}
\left.\eta_T=\pi\frac{d\rho^{(s)}_T}{d\omega}\right|_{\omega=0}~~~ {\rm and}~~~
\left.\zeta_T=\frac{\pi}{9}\frac{d\rho^{(b)}_T}{d\omega}\right|_{\omega=0}.
\end{equation}
Here, the superscripts $(s)$ and $(b)$ stand respectively for ``shear'' and ``bulk''.
Both spectral densities can be obtained from the integral equation, called Kubo formula:
\begin{equation}
\label{kubo} 
\int_0^\infty d\omega{\,} \rho_T^{(s),(b)}(\omega){\,} \frac{\cosh\left[\omega\left(x_4-\frac{\beta}{2}
\right)\right]}{\sinh(\omega\beta/2)}=\int d^3x\sum\limits_{n=-\infty}^{+\infty}
U_T^{(s),(b)}({\bf x},x_4-\beta n).
\end{equation}
In this equation, $U_T^{(s)}$ and $U_T^{(b)}$ are the following correlation functions 
of the Yang--Mills energy-momentum tensor $\Theta_{\mu\nu}$:
$$U_T^{(s)}({\bf x},x_4-\beta n)=\bigl<\Theta_{12}(0)\Theta_{12}({\bf x},x_4-\beta n)\bigr>_T,~~~ {\rm where}~~~
\Theta_{12}=g^2F_{1\mu}^aF_{2\mu}^a;$$
\begin{equation}
\label{sp87}
U_T^{(b)}({\bf x},x_4-\beta n)=\bigl<\Theta_{\mu\mu}(0)\Theta_{\nu\nu}({\bf x},x_4-\beta n)\bigr>_T,~~~~ 
{\rm where}~~~~ \Theta_{\mu\mu}=\frac{\beta(g)}{2g}(F_{\mu\nu}^a)^2.
\end{equation}
In Ref.~\cite{1}, Eq.~(\ref{kubo}) for $\rho_T^{(s)}(\omega)$
was solved by means of the Fourier transform. Owing to the uniformity of Eq.~(\ref{kubo}), this 
approach equally applies to $\rho_T^{(b)}(\omega)$. Its main idea is to use for the 
nonperturbative parts of the zero-temperature correlation functions
$U_{T=0}^{(s),(b)}(x)$ the following ansatz, exponentially falling off with distance:
\begin{equation}
\label{cor0}
U_{T=0}^{(s),(b)}(x)=N_\alpha^{(s),(b)}\bigl<G^2\bigr>^2\cdot\frac{K_{2-\alpha}(M|x|)}{(M|x|)^{2-\alpha}}.
\end{equation}
Here $\bigl<G^2\bigr>\equiv\bigl<g^2(F_{\mu\nu}^a)^2\bigr>$ is the gluon condensate,  
$K_{2-\alpha}$ is the MacDonald function, and $\alpha\in(0,\infty)$ is some parameter. 
Assuming parametrization~(\ref{cor0}), one gets at $T>T_c$ the following
Fourier transform of Eq.~(\ref{kubo}):
\begin{equation}
\label{om12}
\int_0^\infty d\omega{\,} \rho_T^{(s),(b)}(\omega){\,}
\frac{\omega}{\omega^2+\omega_k^2}=\pi^2 2^\alpha\Gamma(\alpha)N_\alpha^{(s),(b)}
\left<G^2\right>_T^2
\frac{M_T^{2\alpha-4}}{(\omega_k^2+M_T^2)^\alpha},
\end{equation}
where $\Gamma(\alpha)$ is the Gamma-function. Note that, while the gluon condensate and the correlation
length $1/M$ become temperature-dependent, the overall coefficients $N_\alpha^{(s),(b)}$ are determined entirely by the corresponding zero-temperature correlation functions $U_{T=0}^{(s),(b)}(x)$. Next, one solves Eq.~(\ref{om12}) by imposing for $\rho_T^{(s),(b)}(\omega)$ a Lorentzian-type ansatz~\cite{kw, kt}
\begin{equation}
\label{param1}
\rho_T^{(s),(b)}(\omega)=C_T^{(s),(b)}\cdot\frac{\omega}{(\omega^2+M_T^2)^{\alpha+\frac12}}.
\end{equation}
For any $\alpha\in (0,\infty)$, it provides convergence of the $\omega$-integration in Eq.~(\ref{om12}), and 
guarantees that both sides of Eq.~(\ref{om12}) have the same large-$|k|$ behavior. Furthermore, it turns 
out that only for a single value of $\alpha$, namely $\alpha=1/2$, the spectral densities, once 
sought in the form of Eq.~(\ref{param1}), appear $k$-independent~\cite{1}. [As can be seen from Eq.~(\ref{param1}), at $\alpha=1/2$, the spectral densities take the purely Lorentzian form considered in Refs.~\cite{kw, kt}.] 
The corresponding temperature-dependent functions $C_T^{(s),(b)}$ can then be found, and read
\begin{equation}
\label{cT}
C_T^{(s),(b)}=(2\pi)^{3/2}N_{1/2}^{(s),(b)}\cdot\frac{\bigl<G^2\bigr>_T^2}{M_T^3}.
\end{equation}

Thus, to find the viscosities, it remains to determine the coefficients $N_{1/2}^{(s),(b)}$.
To this end, one evaluates the correlation functions
$U_{T=0}^{(s)}(x)=\bigl<\Theta_{12}(0)\Theta_{12}(x)\bigr>$ and 
$U_{T=0}^{(b)}(x)=\bigl<\Theta_{\mu\mu}(0)\Theta_{\nu\nu}(x)\bigr>$
via the so-called Gaussian-dominance hypothesis.  
This hypothesis, supported by the lattice simulations~\cite{dmp, na, latt}, 
states that the connected four-point correlation function of gluonic field strengths
can be neglected compared to pairwise products of the two-point correlation functions.
For the correlation function 
$U_{T=0}^{(s)}(x)=\bigl<g^4F_{1\mu}^a(0)F_{2\mu}^a(0)F_{1\nu}^b(x)F_{2\nu}^b(x)\bigr>$,
this approximation yields
$$U_{T=0}^{(s)}(x)\simeq
\left<g^2F_{1\mu}^a(0)F_{2\mu}^a(0)\right>\left<g^2F_{1\nu}^b(x)F_{2\nu}^b(x)\right>+
\left<g^2F_{1\mu}^a(0)F_{1\nu}^b(x)\right>\left<g^2F_{2\mu}^a(0)F_{2\nu}^b(x)\right>+$$
\begin{equation}
\label{uj}
+\left<g^2F_{1\mu}^a(0)F_{2\nu}^b(x)\right>\left<g^2F_{2\mu}^a(0)F_{1\nu}^b(x)\right>.
\end{equation}
The sought nonperturbative contributions to the viscosities, produced by the background fields, can now be parametrized by means of the stochastic vacuum model~\cite{ds}. In this section, we consider 
only the contribution produced by {\it confining} self-interactions of the background fields, while the 
contribution of {\it nonconfining} nonperturbative self-interactions will be considered in the next section.
Confining self-interactions of the background fields enter the two-point correlation function of 
gluonic field strengths as~\cite{ds, mpla}
\begin{equation}
\label{sta}
\bigl<g^2F_{\mu\nu}^a(x)F_{\lambda\rho}^b(0)\bigr>=\frac{\bigl<G^2\bigr>}{12}\cdot
(\delta_{\mu\lambda}\delta_{\nu\rho}-
\delta_{\mu\rho}\delta_{\nu\lambda})\cdot\frac{\delta^{ab}}{N_c^2-1}\cdot D(x).
\end{equation}
For the rest of this section, we set $N_c=3$. At large $|x|$, the dimensionless function $D(x)$ falls off 
as ${\rm e}^{-\mu|x|}$, where $1/\mu$ is the so-called vacuum correlation length~\cite{ds, dmp, na} 
(for a review see~\cite{latt}).
The compatibility of Eqs.~(\ref{uj}) and (\ref{sta})
with Eq.~(\ref{cor0}), at $\alpha=1/2$, is achieved by having $M=2\mu$ and 
\begin{equation}
\label{d12}
D(x)={\cal A}\cdot\sqrt{\frac{K_{3/2}(2\mu|x|)}{(2\mu|x|)^{3/2}}}.
\end{equation}
Plugging Eqs.~(\ref{sta}) and (\ref{d12}) into Eq.~(\ref{uj}), one obtains the desired coefficient 
$N_{1/2}^{(s)}$ in terms of the constant ${\cal A}$:
\begin{equation}
\label{nA}
N_{1/2}^{(s)}=\frac{{\cal A}^2}{576}.
\end{equation}
The constant ${\cal A}$ can be fixed by 
means of the expression for the string tension in the 
fundamental representation~\cite{mpla}
\begin{equation}
\label{rel88}
\sigma_{\rm f}=\frac{\left<G^2\right>}{144}\int d^2x D(x), 
\end{equation}
which yields
\begin{equation}
\label{calA}
{\cal A}=\frac{4}{\int_0^\infty dz\cdot z^{1/4}\cdot\sqrt{K_{3/2}(z)}}\simeq1.05.
\end{equation}
Finally, substituting Eq.~(\ref{nA}) into Eq.~(\ref{cT}), and using Eqs.~(\ref{vb53}) and (\ref{param1}), one 
accomplishes the calculation of the shear viscosity~\cite{1}:
\begin{equation}
\label{sh54}
\eta_T=\frac{\pi^{5/2}{\cal A}^2}{4608\sqrt{2}}\cdot\frac{\bigl<G^2\bigr>_T^2}{\mu_T^5}.
\end{equation}

A contribution of confining self-interactions of the background fields to the bulk viscosity can be 
found in a similar way. It amounts to calculating the coefficient $N_{1/2}^{(b)}$ in Eq.~(\ref{cor0}). That can be done by using in Eq.~(\ref{sp87}) the one-loop Yang--Mills $\beta$-function, in which case 
\begin{equation}
\label{nv69}
\frac{\beta(g)}{2g}\simeq-\frac{11}{32\pi^2}g^2,~~~ {\rm and}~~~ 
U_{T=0}^{(b)}(x)\simeq\left(\frac{11}{32\pi^2}\right)^2\cdot
\left<g^4F_{\mu\nu}^a(0)F_{\mu\nu}^a(0)F_{\lambda\rho}^b(x)F_{\lambda\rho}^b(x)\right>.
\end{equation}
Furthermore, by means of the Gaussian-dominance hypothesis, the correlation function $U_{T=0}^{(b)}(x)$
can be written as 
\begin{equation}
\label{uB}
U_{T=0}^{(b)}(x)\simeq\left(\frac{11}{32\pi^2}\right)^2\cdot\left[
\left<G^2\right>^2+2\left<g^2F_{\mu\nu}^a(0)F_{\lambda\rho}^b(x)\right>^2\right].
\end{equation}
Applying the parametrization of the stochastic vacuum model, Eq.~(\ref{sta}), we have 
\begin{equation}
\label{gg65}
U_{T=0}^{(b)}(x)\simeq\left(\frac{11}{32\pi^2}\right)^2\left<G^2\right>^2\left[1+\frac{1}{24}D^2(x)\right].
\end{equation}
The renormalized spectral density corresponds to $U_{T=0}^{(b)}(x)$ with the ``1'' in the 
brackets of Eq.~(\ref{gg65}) subtracted. The function $D(x)$ in the form of 
Eq.~(\ref{d12}) yields then the coefficient which enters Eq.~(\ref{cor0}):
$$N_{1/2}^{(b)}=\frac{{\cal A}^2}{24}\left(\frac{11}{32\pi^2}\right)^2.$$
Plugging this coefficient into Eq.~(\ref{cT}), and using Eqs.~(\ref{vb53}) and (\ref{param1}), we 
arrive at the following formula for the bulk viscosity:
\begin{equation}
\label{bu76}
\zeta_T=\frac{{\cal A}^2}{1728\sqrt{2\pi^3}}\left(\frac{11}{32}\right)^2
\cdot\frac{\bigl<G^2\bigr>_T^2}{\mu_T^5}.
\end{equation}
Its numerical evaluation will be performed in section~IV.

\section{Contributions of nonconfining nonperturbative self-interactions of the background fields}

\noindent
The two-point correlation function $\left<g^2F_{\mu\nu}^a(0)F_{\lambda\rho}^b(x)\right>$ 
parametrizes not only confining self-interactions of stochastic background fields, through 
the function $D(x)$, 
but also nonconfining nonperturbative self-interactions, through the so-called function 
$D_1(x)$~\cite{ds, mpla}. The full 
two-point correlation function reads [cf. Eq.~(\ref{sta})]
$$\left<g^2F_{\mu\nu}^a(0)F_{\lambda\rho}^b(x)\right>=\frac{\left<G^2\right>}{12}\cdot
\frac{\delta^{ab}}{N_c^2-1}\cdot
\bigl\{\kappa(\delta_{\mu\lambda}\delta_{\nu\rho}-\delta_{\mu\rho}\delta_{\nu\lambda})D(x)+$$
\begin{equation}
\label{param}
+\frac{1-\kappa}{2}\left[\partial_\mu(x_\lambda\delta_{\nu\rho}-x_\rho\delta_{\nu\lambda})+
\partial_\nu(x_\rho\delta_{\mu\lambda}-x_\lambda\delta_{\mu\rho})\right]D_1(x)\bigr\},
\end{equation}
where $\kappa\in[0,1]$ is some parameter. 

To see that the functions $D(x)$ and $D_1(x)$ indeed parametrize 
respectively the confining and the nonconfining nonperturbative self-interactions, 
one can substitute parametrization~(\ref{param}) into the   
Wilson loop~\footnote{To simplify notations, 
we assume that ``tr'' is normalized by the condition ${\rm tr}{\,}\hat 1=1$, where $\hat 1$ is the unity matrix in the same representation of the SU($N_c$)-group as the generator $T^a$.} $\left<W(C)\right>=
\left<{\rm tr}{\,}P{\,}\exp
\left(ig\oint_{C}^{}dx_\mu T^aA_\mu^a\right)\right>$ expressed through the correlation function $\left<g^2F_{\mu\nu}^a(x)F_{\lambda\rho}^b(x')\right>$ by means of the non-Abelian Stokes' theorem 
and the cumulant expansion~\cite{ds, mpla}. That yields (cf. the second review in Ref.~\cite{ds})
$$\left<W(C)\right>=\exp\biggl\{-\frac{C_2\left<G^2\right>}{96(N_c^2-1)}\biggl[2\kappa\int_{\Sigma_{\rm min}}^{}
d\sigma_{\mu\nu}(x)\int_{\Sigma_{\rm min}}^{}d\sigma_{\mu\nu}(x')D(|x-x'|)+$$
\begin{equation}
\label{w99}
+(1-\kappa)
\oint_{C}^{}dx_\mu \oint_{C}^{}dx'_\mu\int_{(x-x')^2}^{\infty} d\xi D_1(\sqrt{\xi})
\biggr]\biggr\},
\end{equation}
where $C_2$ is the quadratic Casimir operator of a given representation 
(i.e. $T^aT^a=C_2\hat 1$). From this expression, it is explicitly seen that 
the function $D(x)$ mediates confining self-interactions of the background fields, as described by 
the double surface integral, while the function $D_1(x)$ mediates nonconfining nonperturbative self-interactions, as described by the double contour integral.
These two functions can be viewed as phenomenological propagators of the background gluons, 
which describe the two types of self-interactions.
In what follows, we set 
$$D_1(x)=D(x),$$
as suggested by the lattice data~\cite{dmp, na, latt}. This assumption means that, with the separation between two points in Euclidean space,
the nonconfining nonperturbative self-interactions fall off at the same vacuum correlation length $1/\mu$ as the 
confining ones. The two types of self-interactions differ, though, in magnitude, that is 
taken care of by the parameter $\kappa$. In particular, at $\kappa=1$, 
Eq.~(\ref{param}) reproduces Eq.~(\ref{sta}), 
implying the pure area-law of the Wilson loop and the 
full suppression of the nonconfining nonperturbative self-interactions.
Instead, the opposite limit, $\kappa=0$, 
describes a nonconfining vacuum, in which the Wilson loop respects the pure perimeter-law.
The lattice-simulated value of $\kappa$ in SU(3) Yang--Mills theory is~\cite{latt} 
$\kappa=0.83\pm0.03$. Below, by setting 
$\kappa$ to this realistic value,
\begin{equation}
\label{kap}
\kappa\simeq0.83,
\end{equation}
we relax the approximation $\kappa=1$ adopted in the previous section 
and in the earlier works~\cite{1, epj}.
This way, we account for the contributions to the shear and bulk viscosities produced 
by nonconfining nonperturbative self-interactions of the background fields, whose strength thus amounts to 
17\% of the full strength of nonperturbative self-interactions.

We start with the bulk viscosity, and recalculate the correlation function
$U_{T=0}^{(b)}(x)$, 
Eq.~(\ref{uB}), by using for 
$\left<g^2F_{\mu\nu}^a(0)F_{\lambda\rho}^b(x)\right>$ parametrization~(\ref{param}).
We denote the correlation function with $\kappa\ne 1$ as $U_{T=0}^{\kappa (b)}(x)$. 
A straightforward calculation yields for it the following expression [cf. Eq.~(\ref{gg65})]: 
\begin{equation}
\label{kf}
U_{T=0}^{\kappa (b)}(x)\simeq\left(\frac{11}{32\pi^2}\right)^2\left<G^2\right>^2\left\{1+\frac{1}{24}
\left[D^2+\frac{1-\kappa}{2}DD'+\frac{(1-\kappa)^2}{8}(D')^2\right]\right\}, 
\end{equation}
where $D'\equiv|x|\frac{dD}{d|x|}$. We seek the function $D(x)$ in the form generalizing 
Eq.~(\ref{d12}):
\begin{equation}
\label{dK}
D(x)={\cal A}_\kappa\cdot f_\kappa(\mu|x|),~~ {\rm where}~~ {\cal A}_{\kappa=1}={\cal A}~~ {\rm and}~~ 
f_{\kappa=1}(z)=\sqrt{\frac{K_{3/2}(2z)}{(2z)^{3/2}}}.
\end{equation}
Given that $U_{T=0}^{\kappa (b)}(x)$, Eq.~(\ref{kf}), 
contains terms to the order $(1-\kappa)^2$, 
the function $f_\kappa=f_\kappa(\mu|x|)$ can be sought as a power series 
$$f_\kappa=f_{\kappa=1}+(1-\kappa)f^{(1)}+
(1-\kappa)^2f^{(2)}+{\cal O}\left((1-\kappa)^3\right).$$
Furthermore, the Matsubara-mode independence of the spectral density $\rho_T^{(b)}$ is achieved when 
the correlation function~(\ref{kf}) (with ``1'' in the curly brackets subtracted) has the form of Eq.~(\ref{cor0}) with $\alpha=1/2$. The equation representing this condition,
$$f_k^2+\frac{1-\kappa}{2}f_k f_k'+\frac{(1-\kappa)^2}{8}(f_k')^2=\frac{K_{3/2}(2\mu|x|)}{(2\mu|x|)^{3/2}},$$
yields the following functions $f^{(1)}(\mu|x|)$ and $f^{(2)}(\mu|x|)$:
$$f^{(1)}=-\frac14f'_{\kappa=1},~~~~ f^{(2)}=\frac{1}{16}\left[
f''_{\kappa=1}-\frac{(f'_{\kappa=1})^2}{2f_{\kappa=1}}\right],$$
where again $f'\equiv|x|\frac{df}{d|x|}$. Explicitly, we find 
$$f_\kappa(z)=\frac{\pi^{1/4}}{256\cdot 2^{3/4}}\cdot\frac{{\rm e}^{-z}}{[z(1+2z)]^{3/2}}\cdot\bigl\{
128(1+2z)^2+$$
\begin{equation}
\label{fk7}
+(1-\kappa)\cdot 16(1+2z)(3+6z+4z^2)+(1-\kappa)^2\cdot\left[9+4z\cdot(9+z\cdot(7+4z(1+z)))\right]
\bigr\},
\end{equation}
that reproduces correctly the function 
$$f_{k=1}(z)=\frac{\pi^{1/4}}{2^{7/4}}\cdot\frac{{\rm e}^{-z}}{z^{3/2}}\cdot(1+2z)^{1/2},$$
given by Eq.~(\ref{dK}). Next, much as the coefficient ${\cal A}$, 
the coefficient ${\cal A}_\kappa$ can be found by means of 
relation~(\ref{rel88}), and reads
\begin{equation}
\label{aK}
{\cal A}_\kappa=\frac{1}{\int_0^\infty dz\cdot z\cdot f_\kappa(z)}.
\end{equation}
[In particular, using the explicit form of $f_{\kappa=1}(z)$, Eq.~(\ref{dK}), one can see that 
Eq.~(\ref{aK}) at $\kappa=1$ reproduces Eq.~(\ref{calA}).] Plugging into Eq.~(\ref{aK}) the obtained 
function $f_\kappa$, Eq.~(\ref{fk7}), at the lattice-simulated value of $\kappa$, Eq.~(\ref{kap}), we find
\begin{equation}
\label{aK1}
{\cal A}_{\kappa=0.83}\simeq 0.97.
\end{equation}
The corresponding bulk viscosity at this value of $\kappa$,
\begin{equation}
\label{f98}
\zeta_T^{\kappa=0.83}=\frac{({\cal A}_{\kappa=0.83})^2}{1728\sqrt{2\pi^3}}\left(\frac{11}{32}\right)^2
\cdot\frac{\bigl<G^2\bigr>_T^2}{\mu_T^5},
\end{equation}
appears by a factor of 
\begin{equation}
\label{f99}
\left(\frac{{\cal A}_{\kappa=0.83}}{{\cal A}}\right)^2\simeq0.85
\end{equation}
smaller than the approximate bulk viscosity $\zeta_T^{\kappa=1}$ given by Eq.~(\ref{bu76}).

We proceed now to the shear viscosity, and
recalculate the correlation function $U_{T=0}^{(s)}(x)$, Eq.~(\ref{uj}), using for 
$\left<g^2F_{\mu\nu}^a(0)F_{\lambda\rho}^b(x)\right>$ parametrization~(\ref{param}).
Each of the condensates $\left<g^2F_{1\mu}^a(0)F_{2\mu}^a(0)\right>$ and $\left<g^2F_{1\nu}^b(x)F_{2\nu}^b(x)\right>$ on the RHS of Eq.~(\ref{uj}) can only be proportional to 
$\delta_{12}$, and therefore vanish. Upon the use of parametrization~(\ref{param}),
the terms 
$\left<g^2F_{1\mu}^a(0)F_{1\nu}^b(x)\right>\left<g^2F_{2\mu}^a(0)F_{2\nu}^b(x)\right>$ and 
$\left<g^2F_{1\mu}^a(0)F_{2\nu}^b(x)\right>\left<g^2F_{2\mu}^a(0)F_{1\nu}^b(x)\right>$ read
$$\left<g^2F_{1\mu}^a(0)F_{1\nu}^b(x)\right>\left<g^2F_{2\mu}^a(0)F_{2\nu}^b(x)\right>=$$
\begin{equation}
\label{term2}
=\frac{\left<G^2\right>^2}{72(N_c^2-1)}
\left\{D^2+\frac{1-\kappa}{2}DD'+\frac{(1-\kappa)^2}{8}\left[2\left(\frac{x_1x_2}{|x|}\right)^2+x^2-x_1^2-x_2^2
\right]\left(\frac{dD}{d|x|}\right)^2\right\}
\end{equation}
and
\begin{equation}
\label{term3}
\left<g^2F_{1\mu}^a(0)F_{2\nu}^b(x)\right>\left<g^2F_{2\mu}^a(0)F_{1\nu}^b(x)\right>=
\frac{\left<G^2\right>^2}{72(N_c^2-1)}\cdot\frac{(1-\kappa)^2}{4}\cdot \left(\frac{x_1x_2}{|x|}\right)^2
\left(\frac{dD}{d|x|}\right)^2.
\end{equation}
The sum of Eqs.~(\ref{term2}) and (\ref{term3}) yields
$$
U_{T=0}^{(s)}(x)\simeq\frac{\left<G^2\right>^2}{72(N_c^2-1)}\times$$
\begin{equation}
\label{gg54}
\times
\left\{D^2+\frac{1-\kappa}{2}DD'+\frac{(1-\kappa)^2}{8}\left[4\left(\frac{x_1x_2}{|x|}\right)^2+x^2-x_1^2-x_2^2
\right]\left(\frac{dD}{d|x|}\right)^2\right\}.
\end{equation}
Furthermore, since the self-interactions of stochastic background fields we are studying are 
{\it nonperturbative}, it is legitimate to use in Eq.~(\ref{gg54}) the leading large-$|x|$ approximation. In this approximation, the sum 
$(x_1^2+x_2^2)$ can in general be disregarded compared to $x^2$. We should then compare $x^2$  with 
$4(x_1x_2/|x|)^2$, that is the same as to compare $x^2$ with $2|x_1x_2|$. This comparison 
obviously leads to the inequality $(|x_1|-|x_2|)^2+x_3^2+x_4^2>0$, which means that, in the leading large-$|x|$ approximation, the term 
$4(x_1x_2/|x|)^2$ can be disregarded compared to $x^2$. Equation~(\ref{gg54}) then takes the form
$$
U_{T=0}^{(s)}(x)\simeq\frac{\left<G^2\right>^2}{72(N_c^2-1)}
\left[D^2+\frac{1-\kappa}{2}DD'+\frac{(1-\kappa)^2}{8}(D')^2\right].$$ 
Comparing this expression with Eq.~(\ref{kf}), we conclude that 
the function $D(x)$ has the same form of Eq.~(\ref{dK}) as for the bulk viscosity, 
with the function $f_\kappa(x)$ and the coefficient ${\cal A}_\kappa$ given by Eqs.~(\ref{fk7}) and 
(\ref{aK}), respectively. Thus, we 
obtain the following expression for the shear viscosity with the 
nonconfining nonperturbative self-interactions of the background fields taken into account (at $N_c=3$):
\begin{equation}
\label{f100}
\eta_T^{\kappa=0.83}=\frac{\pi^{5/2}
({\cal A}_{\kappa=0.83})^2}{4608\sqrt{2}}\cdot\frac{\bigl<G^2\bigr>_T^2}{\mu_T^5}.
\end{equation}
It is by 15\% [cf. Eq.~(\ref{f99})] 
smaller than Eq.~(\ref{sh54}), where
only confining self-interactions are taken into account~\cite{1}. In the next section, we proceed with the 
numerical evaluation of the ratios to the entropy density of the bulk and shear viscosities, respectively 
Eqs.~(\ref{bu76}), (\ref{f98}), and (\ref{f100}).

\section{Numerical evaluation}
For the calculation of the bulk viscosity, we use the one-loop running coupling [cf. Eq.~(\ref{nv69})]
$$g^{-2}_T=2b_0\ln\frac{T}{\Lambda},~~~ {\rm where}~~~
b_0=\frac{11N_c}{48\pi^2},$$
and, for the $(N_c=3)$-case under study, $\Lambda=0.104T_c$, $T_c=270{\,}{\rm MeV}$~\cite{f1}.
This coupling should be plugged into the function
\begin{equation}
\label{fT9}
f(T)\equiv \left\{\begin{array}{rcl}1~~ {\rm at}~~ T_c<T<T_{*},\\
\frac{g^2_T\cdot T}{g^2_{T_{*}}\cdot T_{*}}~~ {\rm at}~~ T>T_{*},\end{array}\right.
\end{equation}
which defines the temperature-dependent inverse vacuum correlation length and the 
chromomagnetic gluon condensate as~\cite{yus1, epj, 1}
$$\mu_T=\mu\cdot f(T),~~ \left<G^2\right>_T=\left<G^2\right>\cdot f^4(T).$$
The corresponding zero-temperature values of these quantities are $\mu=894{\,}{\rm MeV}$~\cite{dmp}
and $\left<G^2\right>=\frac{72}{\pi}\sigma_{\rm f}\mu^2\simeq
3.55{\,}{\rm GeV}^4$~\cite{mpla}, where $\sigma_{\rm f}=(440{\,}{\rm MeV})^2$ is the 
string tension in the fundamental representation. Furthermore, the temperature of the dimensional 
reduction $T_{*}$ in Eq.~(\ref{fT9}) can be obtained from the equation
$\sigma_{\rm f}(T_{*})=\sigma_{\rm f}$,  
where $\sigma_{\rm f}(T)=[0.566g^2_T\cdot T]^2$ is 
the high-temperature parametrization of the spatial string tension~\cite{f1}.  
Solving this equation numerically, one gets~\cite{1}
$$T_{*}=1.28T_c.$$
Finally, from the lattice data for the pressure of the gluon plasma~\cite{f1}
$p_{\rm lat}=p_{\rm lat}(T)$, one obtains the entropy density $s=dp_{\rm lat}/dT$.
Its plot can be found in Ref.~\cite{1}.

\begin{figure}
\psfrag{G}{\Large{$\zeta_T/s$}}
\epsfig{file=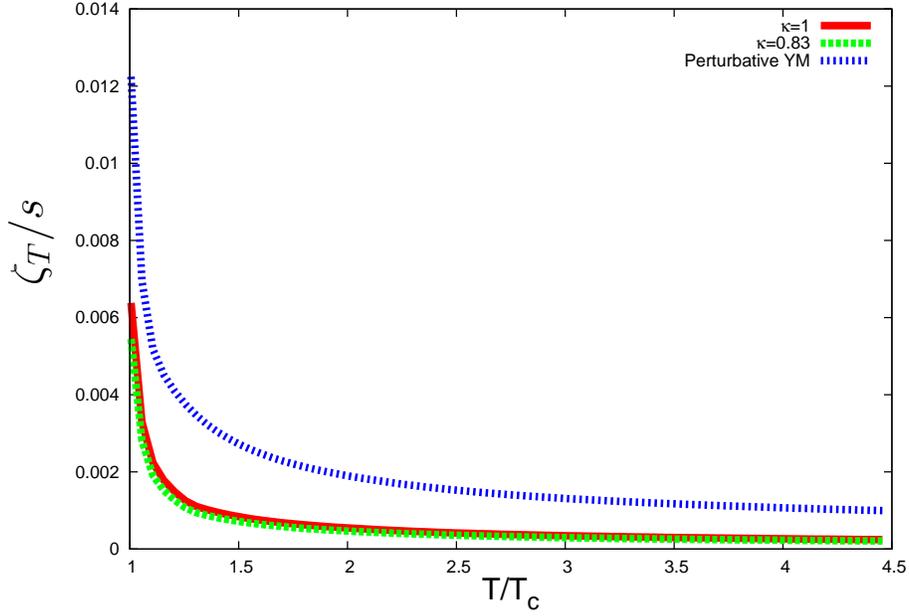, width=120mm}
\caption{\label{tt6}Calculated ratios $\zeta_T/s$ and $\zeta_T^{\kappa=0.83}/s$
as functions of temperature. Also shown for comparison are perturbative values $\zeta_{\rm pert}/s$, where $\zeta_{\rm pert}$ is given by Eq.~(\ref{pertZ}) 
and extrapolated down to $T=T_c$.}
\end{figure}

With the temperature dependence of all the quantities fixed in this way, we numerically 
calculate the ratios to the entropy density of the bulk viscosities given by Eqs.~(\ref{bu76}) and (\ref{f98}). The 
results of this calculation are plotted in Fig.~\ref{tt6}.
For comparison, in the same Fig.~\ref{tt6}, we plot the ratio $\zeta_{\rm pert}/s$, where the perturbative bulk viscosity
\begin{equation}
\label{pertZ}
\zeta_{\rm pert}=\frac{0.443\alpha_s^2T^3}{\ln(7.14/g_T)},
\end{equation}
with $\alpha_s\equiv g_T^2/(4\pi)$,
was obtained in Ref.~\cite{pertZeta} in the leading logarithmic approximation. 
For illustrative purposes, we 
extrapolate this weak-coupling formula down to $T=T_c$.
At temperatures $T\gtrsim 2T_c$, our results scale as $\frac{\zeta_T}{s}\sim \frac{\zeta_T^{\kappa=0.83}}{s}\sim
g_T^6$ (cf. Introduction), whereas the perturbative result~(\ref{pertZ}) scales as $\frac{\zeta_{\rm pert}}{s}\sim g_T^4$. One observes a qualitative agreement 
between these two scaling laws and the corresponding curves in Fig.~\ref{tt6}.

\begin{figure}
\psfrag{S}{\Large{$\eta_T/s$}}
\epsfig{file=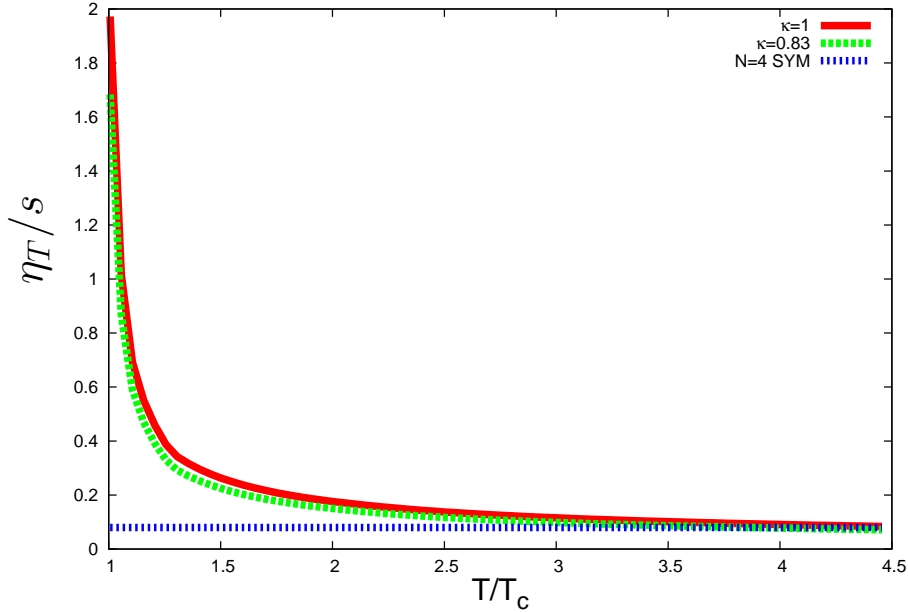, width=120mm}
\caption{\label{tt64}Calculated ratio $\eta_T^{\kappa=0.83}/s$ as a function of temperature. 
Also shown for comparison are $\eta_T/s$, Eq.~(\ref{sh54}), and the 
conjectured lower bound of $1/(4\pi)$ realized in ${\cal N}=4$ SYM.}
\end{figure}

For the calculation of the $(\eta_T^{\kappa=0.83}/s)$-ratio, where $\eta_T^{\kappa=0.83}$ is given by Eq.~(\ref{f100}), we use the two-loop running coupling~\cite{f1}
$$g^{-2}_T=2b_0\ln\frac{T}{\Lambda}+\frac{b_1}{b_0}\ln\left(2\ln\frac{T}{\Lambda}\right),~ {\rm where}~
b_1=\frac{34}{3}\left(\frac{N_c}{16\pi^2}\right)^2.$$
The corresponding numerical results as a function of temperature are plotted in Fig.~\ref{tt64}.
The $(\eta_T/s)$-ratio, with $\eta_T$ given by Eq.~(\ref{sh54}), 
is also shown in Fig.~\ref{tt64} for comparison. The  
numerical values of the $(\eta_T^{\kappa=0.83}/s)$-ratio 
exhibit only a small decrease due to the nonconfining 
nonperturbative interactions. At temperatures $T\gtrsim 2T_c$, where
$\frac{\eta_T}{s}\sim\frac{\eta_T^{\kappa=0.83}}{s}\sim g_T^6$, the calculated contributions of  
stochastic background fields to the shear-viscosity 
to the entropy-density ratio become subdominant compared to the 
contribution of valence gluons. The latter should gradually provide an increase of the full 
shear-viscosity to the entropy-density ratio towards the perturbative result, which is of the order of 
${\cal O}\left(\frac{1}{g_T^4\ln\frac{\rm const}{g_T}}\right)$. The same applies to the 
bulk-viscosity to the entropy-density ratio, whose calculated nonperturbative ${\cal O}(g_T^6)$-part
at $T\gtrsim 2T_c$ should be gradually taken over by the perturbative 
${\cal O}\left(\frac{g_T^4}{\ln\frac{\rm const}{g_T}}\right)$-contribution of valence gluons.

\section{Concluding remarks}
In this paper, we calculated the contributions of stochastic background fields to the 
shear and bulk viscosities of the gluon plasma in SU(3) Yang--Mills theory. These contributions
correspond to two types of nonperturbative self-interactions of the background fields, namely the 
confining and the nonconfining ones. While the contribution of confining self-interactions to the 
shear viscosity had already been obtained in Ref.~\cite{1}, here we calculated the contributions of 
both types of self-interactions to the both viscosities. Our method is based on the Kubo formulae,
by means of which the correlation functions of the energy-momentum tensor, receiving the two above-mentioned types of nonperturbative contributions, can yield the spectral functions of the shear and bulk viscosities. The condition of the Matsubara-mode independence of 
the spectral functions, together with their Lorentzian shape~\cite{kw, kt}, leads to the unique correlation function of gluonic field strengths, given by
Eqs.~(\ref{dK})-(\ref{aK1}). Remarkably, this correlation function is the same for both the bulk
and the shear viscosities. Its amplitude, Eq.~(\ref{aK1}), defines the two viscosities with the 
nonconfining nonperturbative self-interactions of stochastic background fields taken into account
[cf. Eqs.~(\ref{f98}) and (\ref{f100})]. Numerical results for the ratios of the viscosities to the 
entropy density are plotted in Figs.~\ref{tt6} and \ref{tt64}. They show that the found 15\%-decrease 
of the viscosities due to the nonconfining nonperturbative self-interactions 
does not lead to somewhat significant deviations of the two ratios from their values corresponding to
the confining self-interactions alone. The amount of this decrease is close to 17\%, that is the 
relative contribution of the nonconfining part to the nonperturbative self-interactions 
of the background fields.

At sufficiently high temperatures, $T\gg T_c$, contributions to the viscosities produced by valence gluons 
should dominate over the above-calculated contributions of stochastic background fields.
In particular, the purely perturbative contributions are strictly additive to those  
of the background fields~\cite{1}. Instead, at smaller temperatures, 
$T\sim T_c$, a mixing can occur between the nonperturbative contributions of the background fields and 
those of the {\it spatially confined} valence gluons. In Ref.~\cite{yus1}, such a mixing was studied
for the pressure and the interaction measure of the gluon plasma. It was shown there that, for these 
thermodynamic quantities, spatial confinement of valence gluons plays a small role
at temperatures $T_c<T<T_{*}$ compared to other nonperturbative effects. For this reason, one can expect that 
its role is small for the transport coefficients as well. In particular, it is unlikely that the valence 
gluons can somewhat significantly change the obtained 
rapid decrease of the background-fields' contribution to the 
shear-viscosity to the entropy-density ratio at $T_c<T<T_{*}$. The change of this decrease to the 
${\cal O}\left(\frac{1}{g_T^4\ln\frac{\rm const}{g_T}}\right)$-increase 
at $T\gg T_c$ implies the existence of a minimum of the full shear-viscosity to the entropy-density ratio 
at intermediate temperatures.
Once found, the temperature at which this minimum occurs 
can be associated with a transition of the gluon plasma 
from the phase of a strongly interacting quantum liquid to the phase of a weakly interacting 
gas of gluons (cf. the minima of the shear-viscosity to the entropy-density ratio occuring 
nearby liquid-gas phase transitions for water, helium, and nitrogen, mentioned in Introduction). 
In the forthcoming publications, we plan to quantify these statements by 
explicitly calculating contributions of valence gluons to the shear and bulk viscosities.

\acknowledgments

\noindent
This work was started during the author's stay at the University of Bielefeld under the 
support by the German Research Foundation (DFG), contract Sh~92/2-1. At the final stage, 
the work was supported by the 
Centre for Physics of Fundamental Interactions (CFIF) at Instituto Superior T\'ecnico (IST), Lisbon.
The author is grateful to J.E.F.T.~Ribeiro, O.~Kaczmarek, F.~Karsch, E.~Meggiolaro, and A.~Shoshi for the useful discussions.  He also thanks  
F.~Karsch for providing the details of the lattice data from Ref.~\cite{f1}.

\end{document}